\documentclass[prb,floatfix,twocolumn,showpacs,aps]{revtex4}
\usepackage{graphicx}
\usepackage{dcolumn}
\usepackage{amsmath}
\usepackage{natbib}
\usepackage{amsfonts}
\usepackage{amsmath}
\usepackage{amssymb}
\usepackage{graphicx}

\input{tcilatex}

\begin{document}

\author{A.L. Cornelius and B.E. Light}
\affiliation{Department of Physics, University of Nevada, Las Vegas, Nevada, 89154-4002}
\author{J.J. Neumeier}
\affiliation{Department of Physics, Montana State University, Bozeman, MT 59717}
\title{Evolution of the Magnetic Ground State in the Electron-Doped
Antiferromagnet CaMnO$_{3}$}
\date{\today }

\begin{abstract}
Measurements of the specific heat on the system Ca$_{1-x}$La$_{x}$MnO$_{3}$ (%
$x\leq 0.10$) are reported. Particular attention is paid to the effect that
doping the parent compound with electrons by substitution of La for Ca has
on the magnetic ground state. The high ($T>40$ K) temperature data reveals
that doping decreases $T_{N}$ from 122 K for the undoped sample to 103 K for 
$x=0.10$. The low temperature ($T<20$ K) heat capacity data is consistent
with phase separation. The undoped sample displays a finite density of
states and typical antiferromagnetic behavior. The addition of electrons in
the $x\leq 0.03$ samples creates local ferromagnetism as evidenced by a
decreased internal field and the need to add a ferromagnetic component to
the heat capacity data for $x=0.03$. Further substitution enhances the
ferromagnetism as evidenced by the formation of a long range ferromagnetic
component to the undoped antiferromagnetic structure. The results are
consistent with a scenario involving the formation of isolated ferromagnetic
droplets for small $x$ that start to overlap for $x\approx 0.06$ giving rise
to long range ferromagnetism coexisting with antiferromagnetism.

\end{abstract}

\maketitle

\section{Introduction}

Electronic phase segregation in transition metal oxide systems has become an
important topic in condensed matter physics.\cite%
{Dagotto01,Moreo99,Rao97,Levi98,Tranquada95} Systems which exhibit colossal
magnetoresistance (CMR), high temperature superconductivity, and other
compounds such as the nickelates,\cite{Neumeier01} phase segregate into
electron-rich and electron-poor regions. These phase segregated regions are
known to order and form a stripe phase,\cite{Tranquada95} or the atomic
orbitals of the transition metal ions themselves \cite{Chen97} sometimes
order. Theoretical descriptions of the large magnetoresistance observed in
CMR oxides is currently focused on numerous approaches, but the
phase-segregation scenario, where charge inhomogeneous regions compete with
ferromagnetism,\cite{Dagotto01,Moreo99,Rao97} is currently receiving a great
deal of attention. In this picture, the application of magnetic field, or
decreasing the temperature, alters the ratio of these phase segregated
regions thereby affecting the electrical conductivity. In the case of high
temperature superconductors, the possible relation between the stripe phase
and superconductivity is under debate, but some believe that the stripes are
crucial to the formation of a superconducting ground state in the cuprates.%
\cite{Service99,Emery99}

Some recent studies on Ca$_{1-x}R_{x}$MnO$_{3}$ revealed that doping Ca$%
^{2+} $ with a trivalent ion $R^{3+}$ ($R$ = La, Pr, Sm, Gd, Eu, Ho, or Bi)
leads to the formation of a small ferromagnetic (FM) moment.\cite%
{Chiba97,Troyanchuk97,Maignan98,Maignan98_2} This effect was subsequently
investigated in greater detail \cite{Neumeier00} where the FM moment at 5 K $%
M_{sat}$(5 K) was found to display two distinct regimes for La dopings $%
x<0.08$. In the range $0\leq x\leq 0.02$, the doped electrons appear to
remain essentially localized on Mn sites leading to isolated Mn$^{3+}$ ions
which coexist with the majority Mn$^{4+}$ ions. This leads to a
ferromagnetic moment, shown through a phenomenological model, consistent
with the existence of local ferrimagnetism. That is, each doped electron
creates one local ferrimagnetic site. Doping in the range $0.03\leq x\leq
0.07$ leads to improved electrical conductivity and a FM moment which has a
stronger $x$ dependence. The observed moment is consistent with the
existence of local ferromagnetic regions. This picture of local
ferromagnetic regions coexisting within an antiferromagnetic (AFM)
background is supported by recent NMR experiments,\cite{Savosta00}
theoretical treatments,\cite{Chen01} and neutron powder diffraction
experiments.\cite{Ling01,Granado03} Neutron powder diffraction has revealed
that the region $0\leq x\leq 0.07$ indeed contains a mixture of G-type
antiferromagnetism and ferromagnetism which for $x=0.06$ appears to have a
long-range component. A sample of $x=0.02$ reveals, through small angle
neutron scattering (SANS), ferromagnetic clusters of diameter 10 \AA ; these
clusters overlap at larger $x$ values leading to long range ferromagnetism
at $x=0.06$.\cite{Granado03} Concentrations of $x>0.7$ lead to formation of
the C-type antiferromagnetic phase which coexists over a wide temperature
and concentration range with the G-type phase ($0.07\leq x<0.2$).\cite%
{Ling01} These results suggest that the Ca$_{1-x}R_{x}$MnO$_{3}$ system is
ideal for the study of electronic/magnetic phase segregation in the low
doping limit where FM phase segregated regions begin their nucleation within
an AFM host. The present work involves heat capacity studies of some of
these systems.

\section{Experimental Details}

The specimens were synthesized under identical conditions to minimize
variations attributable to chemical defects.\cite{Roosmalen94}
Stoichiometric quantities of (99.99\% purity or better) CaCO$_{3}$, La$_{2}$O%
$_{3}$, and MnO$_{2}$ were weighed and mixed in an agate mortar for 7 min
followed by reaction for 20 h at 1100 $^{\circ }$C. The specimens were
reground for 5 min, reacted for 20 h at 1150 $^{\circ }$C, reground for 5
min, reacted for 20 h at 1250 $^{\circ }$C, reground for 5 min, reacted for
46 h at 1300 $^{\circ }$C, reground for 5 min, pressed into pellets, reacted
for 17 h at 1300 $^{\circ }$C and cooled at 0.4 $^{\circ }$C/min to 30 $%
^{\circ }$C. Powder x-ray diffraction revealed no secondary phases and
iodometric titration, to measure the average Mn valence, indicates the
oxygen content of all specimens falls within the range $3.00\pm 0.01$.
Magnetic measurements were conducted with a SQUID magnetometer. Heat
capacity measurements, using a standard thermal relaxation method, were
performed in a Quantum Design PPMS\ system equipped with a superconducting
magnet capable of generating a 90 kOe magnetic field.

\section{Results and Discussion}

\subsection{Magnetization}

Magnetization as a function of temperature $M(T)$ at 2 kOe is displayed for
three of the Ca$_{1-x}$La$_{x}$MnO$_{3}$ specimens of this study in Fig. \ref%
{magnet}.
\begin{figure}[tbp]
\includegraphics[width=2.7in]{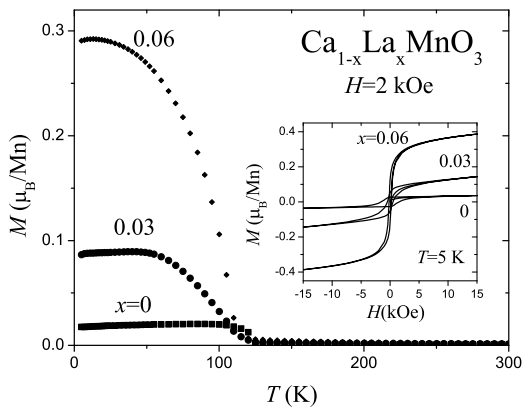}
\caption{ Magnetization $M$ versus temperature in an applied field of 2 kOe
for Ca$_{1-x}$La$_{x}$MnO$_{3}$ specimens. In the inset $M$ versus magnetic
field $H$ at 5 K is displayed.}
\label{magnet}
\end{figure}
The $x=0$ sample, CaMnO$_{3}$, exhibits an antiferromagnetic transition at $%
T_{N}=131$ K. A weak ferromagnetic component is evident in the data, which
is generally attributed to canting of the AFM moments, although recent work 
\cite{Neumeier00} suggests that it arises from a small defect concentration.
Substitution of La for Ca enhances the saturation moment in a systematic
fashion as described previously.\cite{Neumeier00} $M$ versus magnetic field $%
H$ data at 5 K are displayed in the inset of Fig. \ref{magnet}, illustrating
a typical ferromagnetic response. The saturation moment at 5 K can be
extracted from the data in the inset by drawing straight lines through the
two linear portions of the curves; the intersection point of these lines is
defined as $M_{sat}$(5 K).\cite{Neumeier00} As mentioned above, $M_{sat}$
was shown to possess a systematic dependence on $x$ consistent with the
formation of isolated FM regions for small dopings that cross over to AF\
order with a long range FM\ component. This behavior in Ref. 14 revealed
that beyond $x=0.08$ AFM C-type regions began to nucleate, which is
consistent with simple statistical arguments.\cite{Neumeier01_2} These
results are consistent with the scenario of small ferromagnetic clusters
appearing at low concentrations ($0<x\leq 0.3$) that begin to overlap into
long range ferromagnetism at higher doping ($x\approx 0.06$) and to a
coexistence of G- and C-type AF\ with a long range FM\ component for $x>0.08$%
.

\subsection{Low Temperature Specific Heat}

The low temperature specific heat measurements were performed over the
temperature range 0.34 K$<T<20$ K with no applied magnetic field. The total
specific heat can be written as 
\begin{equation}
C_{p}=C_{elec}+C_{mag}+C_{hyp}+C_{lat}  \label{cp eq}
\end{equation}%
where $C_{elec}$ is the electronic contribution, $C_{mag}$ is from the Mn
magnetic moments, $C_{hyp}$ is from the nuclear moment of $^{55}$Mn, and $%
C_{lat}$ is due to the lattice. The electronic contribution is given by $%
\gamma T$, where $\gamma $ is the Sommerfield coefficient which is related
to the density of electronic states at the Fermi energy. $C_{hyp}$ is given
by $A/T^{2}$ where $A$ is related to the internal hyperfine magnetic field
by the relation \cite{Lees99}%
\begin{equation}
A=\frac{R}{3}\QOVERD( ) {I+1}{I}\QOVERD( ) {\mu H_{hyp}}{k_{B}}^{2},
\end{equation}%
where $I$ is the nuclear moment (5/2 for $^{55}$Mn), $\mu $ is the nuclear
magnetic moment (3.45 nuclear magnetons for Mn), and $H_{hyp}$ is the
internal field strength at the Mn site. $C_{lat}$ is estimated by the
approximation $C_{lat}=\beta _{3l}T^{3}$. The Debye temperature $\Theta _{D}$
is given by 
\begin{equation}
\Theta _{D}=\left( \frac{1.944\times 10^{6}\ r}{\beta _{3l}}\right) ^{1/3},
\label{debye}
\end{equation}%
where $\beta _{3l}$ is in mJ/mol K$^{4}$ and $r$ is the number of atoms per
unit cell. Due to the small variation in chemical composition in the
measured samples, it was assumed that $\beta _{3l}$, and therefore $\Theta
_{D}$, are the same for all three samples (this assumption seems valid when
looking at the results of Lees \textit{et al.} who found a variation in $%
\Theta _{D}$ of less than 1.5\% over a much larger range of substitutions 
\cite{Lees99} and our own finding that $\Theta _{D}$ from our high
temperature data is $650\pm 10$ K for all five samples). Like other reports,%
\cite{Lees99,Woodfield97} we added a $T^{5}$ term to adequately fit the
lattice heat capacity data. $C_{mag}$ is estimated as $\sum \beta _{n}T^{n}$
where the value of the exponent $n$ corresponds to the type of magnetic
excitation ($n=3$ for AFM, $n=3/2$ for FM, $n=2$ for a possible
long-wavelength spin excitation \cite{Woodfield97}). For AFM\ and FM\
excitations with a nonzero gap, the heat capacity coefficients are related
to the spin wave stiffness $D$ by the relation $\beta
_{n}=ck_{B}(k_{B}T/D)^{n}$ where $c$ is a constant that depends on the
lattice type.\cite{Lees99} One plausible scenario leading to a $T^{2}$ term
is the \ existence of a long wavelength spin-wave excitation with a planar
FM component of stiffness $D_{\rho }$ and a linear component $D_{z}$ with $%
\beta _{n}\propto (D_{\rho }D_{z})^{-1}.$\cite{Woodfield97}

\begin{figure}[tbp]
\includegraphics[width=2.7in]{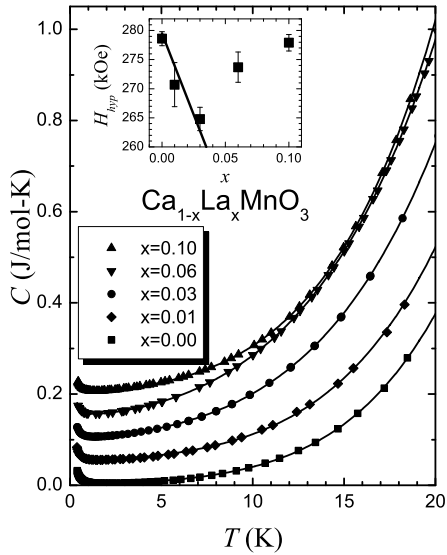}
\caption{Measured specific heat as a function of temperature for {Ca$_{1-x}$%
La$_{x}$MnO$_{3}$}. The lines are fits described in text with the relevant
fitting parameters summarized in Table I. Subsequent curves were offset for
clarity. The inset shows the internal magnetic field $H_{hyp}$ at the Mn
site as deduced from the nuclear contribution to the heat capacity.}
\label{cp low}
\end{figure}

The low temperature measurements are shown in Fig. \ref{cp low}.%
For $x=0$, the data can be fit well using Eq. (\ref{cp eq}) as seen by the
solid line. Since the sample is known to have G-type AFM\ order (each Mn
magnetic moment is aligned antiparallel to its neighbor),\cite{Wollan55} it
is expected that $\beta _{mag}=\beta _{3m}T^{3}.$ Since the lattice term is
also proportional to $T^{3},$ the total $T^{3}$ term $\beta _{3}$ will be $%
\beta _{3l}+\beta _{3m}$, and it is impossible to separate the magnetic and
lattice components, without prior knowledge of the $\Theta _{D}$, from the
low temperature data. However, since the high temperature values of $\Theta
_{D}$ are all similar, any variation we find in $\beta _{3}$ should be
mostly due to $\beta _{3m}$. The parameters which give the best fits to the
data are shown in Table I. 
\begin{table}[tbp]
\caption{Summary of the fitting parameters to the data in Fig.~2.
Definitions of the various coefficients are given in the text. The units are
kOe for $H_{hyp}$, mJ/mol K$^{2}$ for $\protect\gamma $ and mJ/mol K$^{n-1}$
where $n$ is the subscript of the coefficient. The number in parentheses is
the statistical uncertainty in the last digit from the least squares fitting
procedure.}\narrowtext        
\begin{tabular}{llllllll}
& $x$ & $H_{hyp}$ & $\gamma $ & $10^{3}\beta _{3}$ & $10^{5}\beta _{5}$ & $%
\beta _{3/2}$ & $\beta _{2}$ \\ 
\tableline & 0 & 279(1) & 1.38(3) & 2.09(2) & 5.64(4) & - & - \\ 
& 0.01 & 271(3) & 2.44(5) & 3.26(5) & 5.06(9) & - & - \\ 
& 0.03 & 265(2) & 1.46(9) & 4.47(7) & 4.94(9) & 1.17(6) & - \\ 
& 0.06 & 274(2) & 1.98(8) & 3.29(2) & 6.95(9) & - & 0.76(2) \\ 
& 0.10 & 279(2) & 2.76(9) & 5.89(6) & 8.17(7) & 0.35(2) & 
\end{tabular}%
.
\end{table}
From the fit parameters, we can arrive at the following conclusions. The
value of $\beta _{5}$ does not vary much from sample to sample, which gives
us confidence that our fitting parameters are not skewed by the addition of
this term. There exists a nonzero electronic term $\gamma $ indicative of a
finite value of the density of states at the Fermi level, consistent with
thermopower measurements.\cite{Maignan98} Since the data can be fit to a
form $\beta _{n}T^{n}$ leads to the conclusion that there is not a gap in
the spin-wave excitation which often requires an activated term to describe
the heat capacity. The hyperfine term for the undoped sample corresponds to
an internal field of 279 kOe, which is $\sim 30\%$ less than found for LaMnO$%
_{3}$,\cite{Woodfield97} but a factor of 2.6 smaller than determined for Pr$%
_{0.6}$Ca$_{0.4}$MnO$_{3}\,.$\cite{Lees99}

The lines in Fig. 1 are fits to the data using Eq. (\ref{cp eq}) with the
values for $n$ which give the best fits to the data. The fit parameters are
summarized in Table I. The introduction of the electron dopants leads to an
increase in $\gamma $ as one would expect. For $x=0.01$ adding an $n=3/2$ or 
$n=2$ component does not significantly improve the fit. Rather only an
increase in the AF\ term $\beta _{3}$ (note that a larger value of $\beta
_{3}$ relative to the undoped sample corresponds to a smaller spin-wave
stiffness) and a decrease $H_{hyp}$ are found. For $x=0.03$, we find that
there is a need to add a FM $(n=3/2)$ term in addition to a larger AFM $%
(n=3) $ term relative to the $x=0$ value to fit the data. The value of $%
H_{hyp}=265 $ kOe is reduced relative to the undoped sample. One scenario
that would lead to a reduced internal magnetic field would be for the doped
electrons to reside on a Mn site with a spin antiparallel to the original
spin, and larger in magnitude. This would change the local environment
around the\ doped Mn ion to FM while reducing the average AFM magnetic
moment (and thus $H_{hyp}$) and would lead to a local distortion or FM\
polaron. This scenario would lead to a linear decrease in $H_{hyp}$ as $x$
increases. This is exactly what is observed for small values of $x$ in the
inset to Fig. \ref{cp low}, where the line represents $H_{hyp}$ decreasing $%
1.6\%$ per percent of La substitution. The fact that $H_{hyp}$ changes
faster than $x$ is consistent with the moments antiparallel to the original
spin moments being larger in magnitude.

These results are in agreement with magnetization measurements which were
also interpreted in terms of the formation of local FM regions (FM\
polarons) \cite{Neumeier00} and SANS\ measurements that show small
ferromagnetic droplets.\cite{Granado03} In this scenario, one would envision
the weakening of the AFM\ spin-density wave and the formation of FM\
excitations; which is exactly the result of the current measurements. These
results agree with the strong weakening in the AFM interaction recently
deduced from Raman scattering and electron paramagnetic resonance studies.%
\cite{Granado}

As doping is increased above 0.03 we see a definite crossover in all of the
heat capacity data. For $x=0.06,$ the data can no longer be fit by a simple
FM\ term, and in fact the $n=3/2$ component is no longer observed (fitting
with a 3/2 term yields a negative value for $\gamma $). Now an AFM $(n=3)$
term plus a spin density wave $(n=2)$ term are needed. The magnitude of the
AFM\ term decreases and the hyperfine term increases and is nearly the same
as the undoped value. This leads to the conclusion that the nature of the
magnetic excitations has changed and that there might no longer be local FM\
polarons. As mentioned, the $T^{2}$ term can be indicative of a long
wavelength excitation with both FM\ and AFM\ components. These results are
in excellent agreement with the finding of isolated ferromagnetic droplets
in an $x=0.02$ sample which merge into a long range ferromagnetic component
for $x=0.06$.\cite{Granado03} This is consistent with a recent calculation 
\cite{Chen01} predicting a phase transition from the FM polaron state to a
long range FM\ state at a doping of $x\simeq 0.045$ and experimental
findings that there is a crossover from the magnetoelestic polaron to
spin-canted phase around $x=0.06.$\cite{Cohn02} Even higher concentrations ($%
x=0.10$) are again best fit with a FM $n=3/2$ term and a larger $n=3$ term.
This is consistent with neutron scattering results that show the coexistence
of C- and G- type AF\ order at this concentration\cite{Ling01} and
magnetization measurements that show a ferromagnetic moment.

\subsection{High Temperature Specific Heat}

\begin{figure}[tbp]
\includegraphics[width=2.7in]{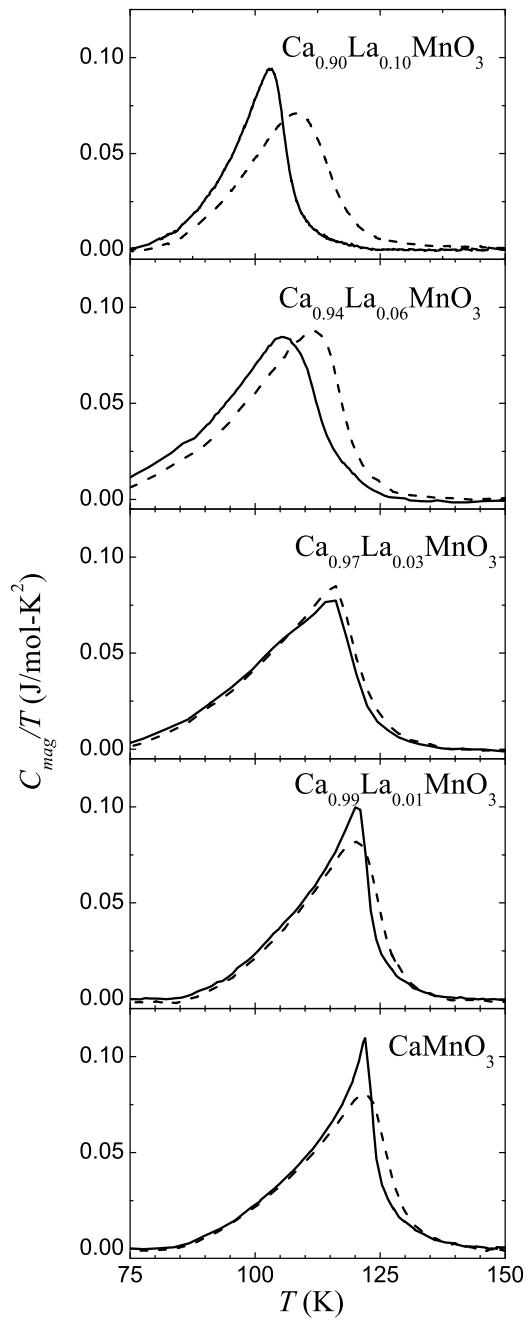}
\caption{Measured magnetic heat capacity $C_{mag}$ as a function of
temperature $T$ in zero applied field (solid lines) and a 90 kOe magnetic
field (dashed lines).}
\label{cp high}
\end{figure}

\begin{figure}[tbp]
\includegraphics[width=2.7in]{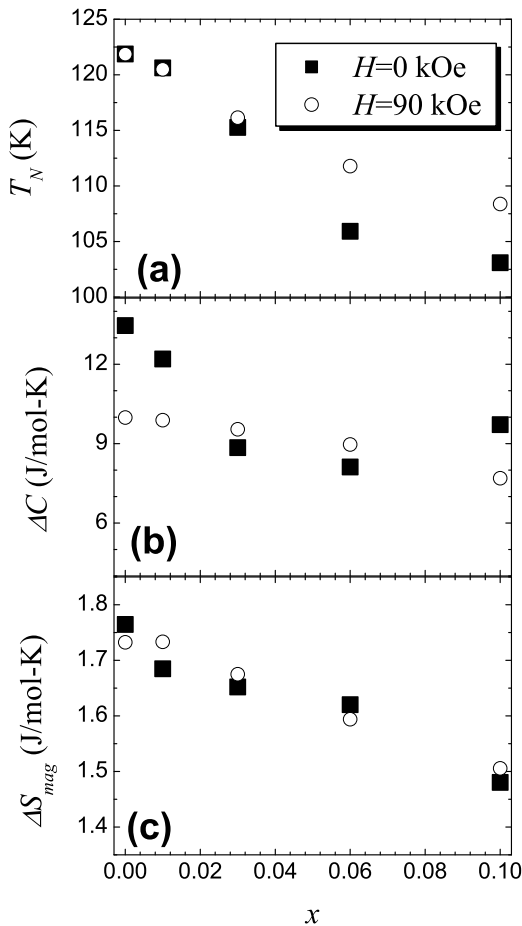}
\caption{Paramters obtained from the data shown in Fig. 3. Solid circles
correspond to the zero field data, while filled circles correspond to data
taken in a 90 kOe applied magnetic field. (a) The magnetic ordering
temperature. (b) The peak value of the magnetic heat capacity $\Delta
C_{mag} $. (c) Change in magnetic entropy $\Delta S_{mag}$, found by
integrating $C_{mag}/T$ as a function of $T$.}
\label{entropy}
\end{figure}

For the high temperature data, $C_{mag}$ is the contribution associated with
the magnetic ordering transitions around $T_{N}$. The high temperature
measurements of $C_{mag}/T$ are shown in Fig. \ref{cp high}.%
In this temperature range $C_{hyp}$ is negligible. The background lattice
contribution was estimated by fitting the measured Debye temperatures above
and below the magnetic transitions with a polynomial. After subtracting this
contribution, we are left with $C_{mag}$. For CaMnO$_{3}$ a magnetic
transition is seen at $T_{N}=122$ K independent of the applied magnetic
field. Doping electrons lowers $T_{N}$ as seen in Fig. \ref{entropy}(a).%
In zero field, the decrease in $T_{N}$ is linear followed by a more rapid
drop for $x>0.3$. In a 90 kOe field, the magnetic ordering temperature obeys
a nearly linear relationship with $x.$ In a similar manner to CaMnO$_{3}$,
the $x=0.01$ and $0.03$ samples do not show a variation in $T_{N}$ as the
magnetic field is increased. This is to be expected as $k_{B}T_{N}\gg \mu
_{B}H$ and the applied field should not alter the antiferromagnetic ordering
temperature. This is not the case for the $x=0.06$ and $x=0.10$ samples
where $T_{N}$ steadily increase in applied field. This increase could be
related to an enhancement of the FM neutron diffraction peak intensity
observed in recent experiments,\cite{Ling01} where at $x=0.12$ FM peak
intensity was seen to increase at the expense of G-type AFM peak intensity
in applied magnetic field. In other words, a long range ferromagnetic
component to the magnetic structure appears to be present for $x=0.06$ and $%
x=0.10$ due to the large enhancement of the magnetic ordering temperature in
an applied filed. The jump in the magnetic heat capacity $\Delta C_{mag}$ at 
$T_{N}$ is shown in Fig. \ref{entropy}(b). The value for the undoped sample
is in good agreement with the work of Moritomo \textit{et al}.\cite%
{Moritomo01} As $x$ increases, $\Delta C_{mag}$ decreases, in sharp contrast
to Moritomo \textit{et al}.\cite{Moritomo01} In a 90 kOe field, the value of 
$\Delta C_{mag}$ is nearly independent of $x.$ To obtain the change in
entropy associated with the magnetic transition $\Delta S_{mag}$ we have
integrated $C_{mag}/T$ versus $T$ and the results are shown in Fig. \ref%
{entropy}(c). Since it is difficult to accurately estimate the lattice
contribution to the heat capacity, the absolute values of $\Delta S_{mag}$
have a reasonable amount of uncertainty. However, changes in $\Delta S_{mag}$
should be much more reliable since the lattice contribution was estimated in
a systematic manner. For all values of $x$ there appears to be little field
dependence to $\Delta S_{mag}.$ For $x=0$ an entropy of $\sim 1.77$ J/mol K
which corresponds to 0.31$R\ln 2$ is found. This value is much smaller than
the value one would expect from Ising spins ($R\ln 2$) or spin 3/2
Heisenberg spins ($R\ln 4$) perhaps due to domain formation. The magnetic
entropy appears to decrease in a nearly linear fashion in a manner similar
to the magnetic ordering temperature.

\section{Conclusion}

We have performed measurements of the specific heat on the system Ca$_{1-x}$%
La$_{x}$MnO$_{3}$ ($x\leq 0.10$) to complement previous magnetization
results. The high ($T>40$ K) temperature data shows that doping decreases
the value of $T_{N}$ from 122 K for the undoped sample to 103 K for $x=0.10$%
. The low temperature ($T<20$ K) heat capacity data is consistent with phase
separation. The undoped ($x=0$) sample displays AFM\ order and has a finite
density of states. When a small amount of holes ($x\leq 0.03)$ are doped
into the sample, local ferromagnetism is introduced into the system. Further
substitution to $x=0.06$ leads to long range ferromagnetism as evidenced by
the formation of a long range spin density wave. These results clearly show
the utility of a thermodynamic measurement (heat capacity) in understanding
the evolution of magnetic properties in electron doped CaMnO$_{3};$ namely
for small dopings small ferromagnetic droplets form which upon further
doping overlap and lead to long range ferromagnetic order coexisting with
antiferromagnetic order.

\end{document}